\documentclass{article}

\usepackage[final]{neurips_2021}

\usepackage[utf8]{inputenc} 
\usepackage[T1]{fontenc}    
\usepackage{hyperref}       
\usepackage{url}            
\usepackage{booktabs}       
\usepackage{amsfonts}       
\usepackage{nicefrac}       
\usepackage{microtype}      
\usepackage{rotating}
\usepackage{forest}

\usepackage{rotating}
\usepackage{graphicx}
\usepackage{amsmath}
\usepackage{amssymb}
\usepackage{amsthm}
\usepackage{dsfont}
\usepackage{multirow}

\usepackage{xcolor}
\usepackage{tikz}
\usetikzlibrary{shapes}
\usetikzlibrary{arrows, arrows.meta}
\usepackage{bm}
\usepackage{dirtytalk}
\usepackage{subcaption}

\definecolor{airforceblue}{rgb}{0.36, 0.54, 0.66}
\definecolor{bleudefrance}{rgb}{0.19, 0.55, 0.91}

\def\@fnsymbol#1{\ensuremath{\ifcase#1\or \dagger\or \ddagger\or
   \mathsection\or \mathparagraph\or \|\or **\or \dagger\dagger
   \or \ddagger\ddagger \else\@ctrerr\fi}}

\title{BRCA Gene Mutations in dbSNP:\\A Visual Exploration of Genetic Variants}

\author{%
Woowon Jang\thanks{Equal contribution, author order is randomized.} \quad Shiwoo Koak$^\ast$ \quad Jiwon Im$^\ast$ \quad Utku Ozbulak\,\thanks{Corresponding author: \texttt{utku.ozbulak@ghent.ac.kr}} \quad Joris Vankerschaver\\
Ghent University, Belgium\\
Ghent University Global Campus, South Korea\\
}

\begin{document}

\maketitle

\begin{figure}[h!]
\centering
\begin{subfigure}{1\textwidth}
\includegraphics[width=0.99\linewidth]{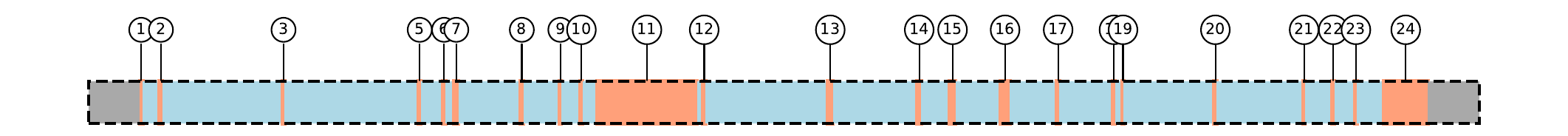}
\caption{\textbf{BR}east \textbf{CA}ncer gene 1}
\end{subfigure}
\begin{subfigure}{1\textwidth}
\includegraphics[width=0.99\linewidth]{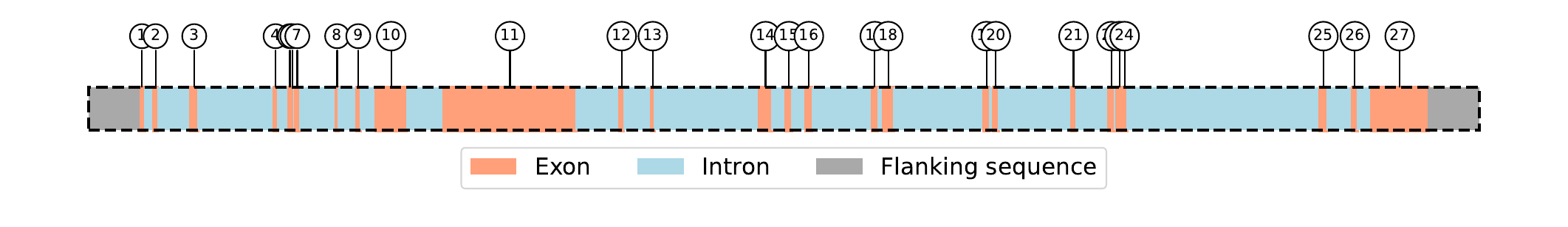}
\caption{\textbf{BR}east \textbf{CA}ncer gene 2}
\end{subfigure}
\caption{Genetic structure visualization of (a) BRCA1 and (b) BRCA2. The length of (orange) exons, (blue) introns, and (gray) flanking sequences are represented according to the length of their base pairs (bp), with intron lengths scaled down by $50\%$ in order to improve visual clarity.}
\label{fig:BRCA_Intron_Exon_visulization}
\end{figure}

\begin{abstract}
BRCA genes, comprising BRCA1 and BRCA2 play indispensable roles in preserving genomic stability and facilitating DNA repair mechanisms. The presence of germline mutations in these genes has been associated with increased susceptibility to various cancers, notably breast and ovarian cancers. Recent advancements in cost-effective sequencing technologies have revolutionized the landscape of cancer genomics, leading to a notable rise in the number of sequenced cancer patient genomes, enabling large-scale computational studies. In this study, we delve into the BRCA mutations in the dbSNP, housing an extensive repository of 41,177 and 44,205 genetic mutations for BRCA1 and BRCA2, respectively. Employing meticulous computational analysis from an umbrella perspective, our research unveils intriguing findings pertaining to a number of critical aspects. Namely, we discover that the majority of BRCA mutations in dbSNP have unknown clinical significance. We find that, although exon 11 for both genes contains the majority of the mutations and may seem as if it is a mutation hot spot, upon analyzing mutations per base pair, we find that all exons exhibit similar levels of mutations. Investigating mutations within introns, while we observe that the recorded mutations are generally uniformly distributed, almost all of the pathogenic mutations in introns are located close to splicing regions (at the beginning or the end). In addition to the findings mentioned earlier, we have also made other discoveries concerning mutation types and the level of confidence in observations within the dbSNP database.
\end{abstract}

\section{Introduction}
Cancer is the most prevalent cause of premature death worldwide, and its dominance continues to increase~\cite{aune2017fruit}. An uncontrolled proliferation of cells is referred to as cancer~\cite{aune2017fruit}. Cancer develops during the cell cycle and uses this mechanism to maliciously spread through tissues by regulating and controlling cell division. Cell cycle checkpoints ensure genetic identity is upheld and eliminate mistakes throughout the cycle. Mutations associated with cancer, however, are capable of continuing cell division without exiting the cell cycle. Continuous uncontrolled cycles give rise to DNA damage and the accumulation of genetic errors, which consequently leads to the development of cancer cells~\cite {matthews2022cell}.

The Global Cancer Observatory examined 36 types of cancer and estimated that, among females, there were 2.3 million new breast cancer cases in 2020~\cite{sung2021global}. Specifically in South Korea, breast cancer was diagnosed with the highest frequency, constituting $30.3\%$ of all cancer types for females between ages 35 to 64 ~\cite{NationalCancerInformationCenter}. Of 60,212 breast cancer cases, a large portion (13,007 cases) are related to genetic mutations in the breast cancer genes (BRCA1 and BRCA2). BRCA1 and BRCA2 are tumor suppressor genes that respond to gene damage during the cell cycle~\cite{roy2012brca1}. BRCA mutations may cause genomic instability, particularly in the context of DNA double-strand breaks (DSBs). DSB refers to the structural alteration of both DNA strands resulting from exposure to ionizing radiation and certain chemicals~\cite{mourad2018predicting,wu2017studies}. BRCA mutations give rise to non-conservative DNA repair and non-homologous end-joining of sister chromatids during various cellular processes, including DSB lesion repair~\cite{kerr2001new}. If the repair of DSB fails, the mutated DNA that is prone to cancer may replicate uncontrollably~\cite{daum2018brca}. 

The detection of genetic mutations is conducted through DNA sequencing~\cite{jiang2013detection}. The introduction of next-generation sequencing (NGS) methods has led to evolutionary improvements in throughput and cost-effectiveness~\cite{miller2022role}. As a result of low-cost sequencing, breast cancer screenings have become more accessible globally, including in developing countries~\cite{melki2023increased}. Therefore, we had access to a variety of public databases on BRCA mutations with their clinical significance. The National Center for Biotechnology Information (NCBI), a public repository for genetic variation, has published the Single Nucleotide Polymorphism database (dbSNP)~\cite{bhagwat2010searching}. By investigating the mutational information of BRCA1 and BRCA2 from dbSNP, we analyze mutations in the exons and introns of the two genes.

\begin{table}[t!]
\centering

\caption{Number of observed mutations from dbSNP for BRCA1 and BRCA2 is categorized according to their position within the genes as exon, intron, or other. The percentage of mutations in each category, relative to the number of mutations in each gene, is presented in parentheses.}
\scriptsize
\begin{tabular}{lrrrr}
\cmidrule[0.25pt]{1-5}
~ & ~ & \multicolumn{3}{c}{\shortstack{Mutation location}} \\
\cmidrule[0.25pt]{3-5}
Gene name & Mutation count & Exon & Intron & Other \\
\cmidrule[0.5pt]{1-5}
BRCA1 & 41,177 & 9,088 ($22.1\%$) & 30,936 ($75.1\%$) & 1,153 ($2.8\%$) \\
BRCA2 & 44,205 & 14,614 ($33.0\%$) & 28,670 ($64.9\%$) & 921 ($2.1\%$) \\
\cmidrule[1pt]{1-5}
\end{tabular}
\label{tbl:total_mutation_count}
\end{table}

\section{Material and Methods}
\label{sec:Material and Methods}

The dbSNP mutation database is a fundamental and comprehensive repository of genetic variations found within the human genome. Maintained and updated by the National Center for Biotechnology Information (NCBI), dbSNP serves as a vital resource for researchers, clinicians, and bioinformaticians in their pursuit of understanding genetic diversity and its implications in health and disease. This valuable database houses a vast collection of single nucleotide polymorphisms, small insertions and deletions, and other genetic variations, along with information on their frequencies in various populations and potential associations with diseases. Researchers use dbSNP to annotate and interpret genetic variants, facilitating genetic association studies, variant prioritization, and the development of personalized medicine approaches. With its constantly expanding dataset, dbSNP continues to play a crucial role in advancing our knowledge of human genetic variation and its impact on human health.

In this study, we utilize BRCA genetic mutations obtained from dbSNP, which houses $41,177$ and $44,205$ records for BRCA1 and BRCA2, respectively. This data obtained from dbSNP is stored in a tab separated file where each row corresponds to a specific genetic mutation. Columns of this data are as follows:
\begin{itemize}
    \item \textbf{Chr}\,\textendash\,Chromosome of the selected gene.
    \item \textbf{Pos}\,\textendash\,Position (of the base pair) in the chromosome of the selected gene where the mutation occurs, based on the GRCh38 reference genome.
    \item \textbf{Variation}\,\textendash\,Detailed description of the mutation provided as the specific change of the affected base pairs. For example, \texttt{A>T} indicates that the base pair at that particular position has changed from \texttt{A} to \texttt{T}.
    \item \textbf{Variant type}\,\textendash\,Signifies the type of the mutation and contains the following values: snv (single nucleotide variation), mnv (multiple nucleotide variation), del (deletion), ins (insertion), and delins (deletion and insertion).
    \item \textbf{Snp id}\,\textendash\,Unique identifier assigned to each genetic variation entry in the database.
    \item \textbf{Clinical significance}\,\textendash\,Significance of the mutation which contains one or more of the following values: benign, likely-benign, likely-pathological, pathological, other, and unknown.
    \item \textbf{Validation status}\,\textendash\,Information regarding the genetic variation on whether it has been experimentally validated or not.
    \item \textbf{Functional class}\,\textendash\,Information about the functional impact of each genetic variation.
    \item \textbf{Gene}\,\textendash\,Gene associated with a specific genetic variant.
    \item \textbf{Frequency}\,\textendash\,Provides frequency or occurrence of a specific genetic variant within different populations.
\end{itemize}

To conduct a comprehensive investigation and facilitate smoother analysis, we have created the following information (i.e., additional columns) using existing data:

\begin{figure}[t!]
\centering
\begin{subfigure}{0.49\textwidth}
\includegraphics[width=1\linewidth]{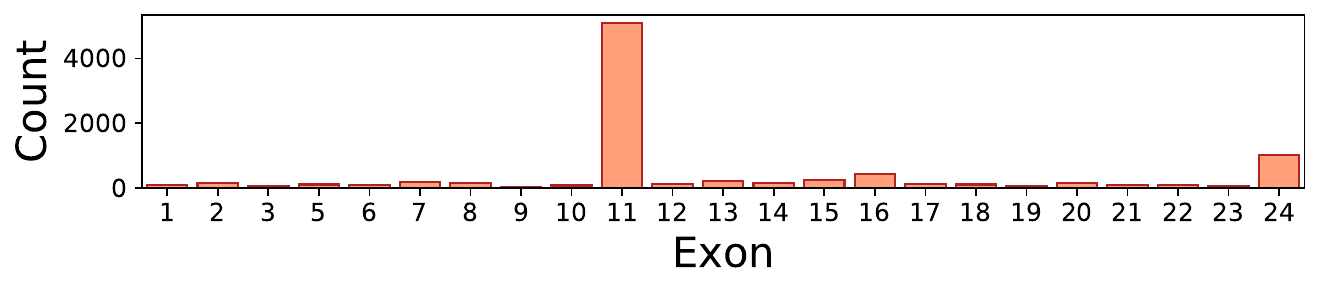}
\includegraphics[width=1\linewidth]{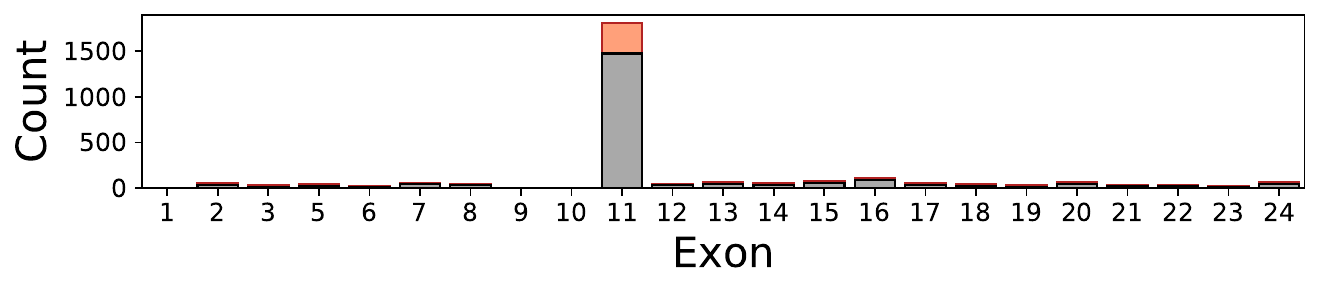}
\caption{BRCA1}
\end{subfigure}
\begin{subfigure}{0.49\textwidth}
\includegraphics[width=1\linewidth]{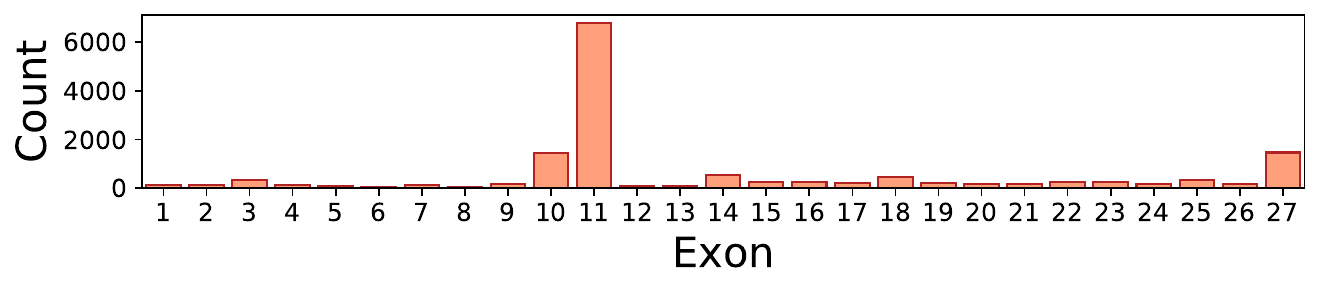}
\includegraphics[width=1\linewidth]{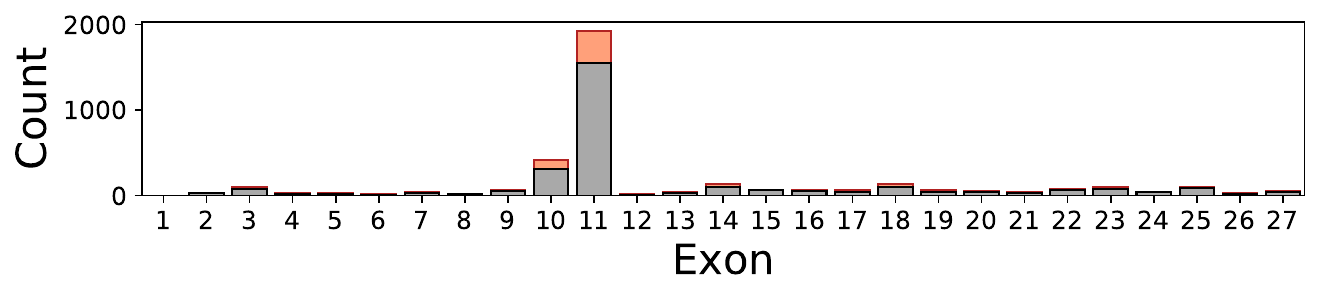}
\caption{BRCA2}
\end{subfigure}
\caption{Distributions of (top) all mutations and (bottom) worst-case pathological mutations across exons of (left) BRCA1 and (right) BRCA2 are provided. In the bottom figures, gray bars represent the amount mutations where best-case and worst-case are both pathological.}
\label{fig:exon_barplot_total_wc_pathogenic_distribution}
\end{figure}

\begin{figure}[t!]
    \centering
\begin{subfigure}{0.49\textwidth}
    \includegraphics[width=1\linewidth]{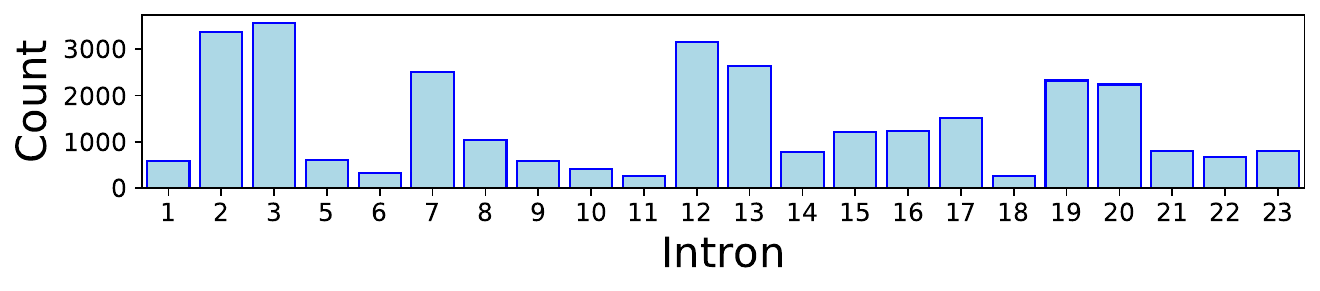}
    \includegraphics[width=1\linewidth]{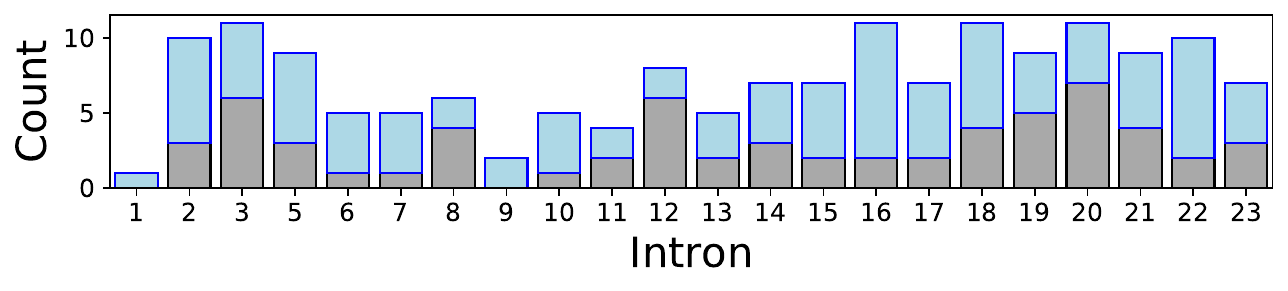}
\caption{BRCA1}
\end{subfigure}
\begin{subfigure}{0.49\textwidth}
    \includegraphics[width=1\linewidth]{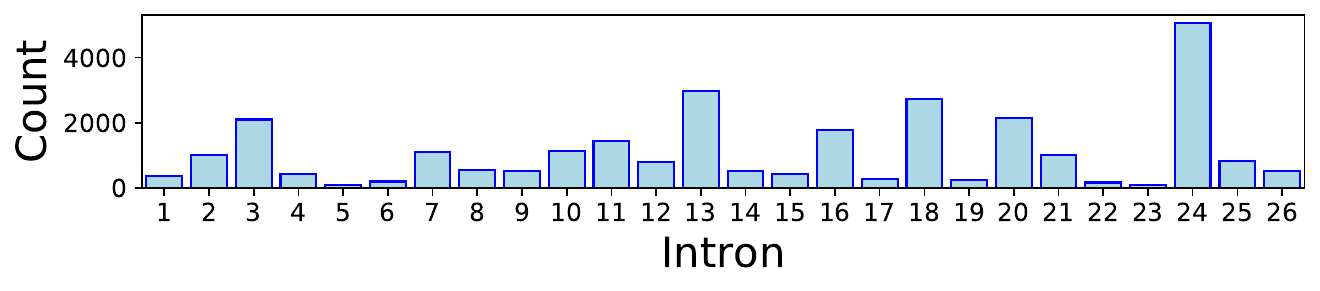}
    \includegraphics[width=1\linewidth]{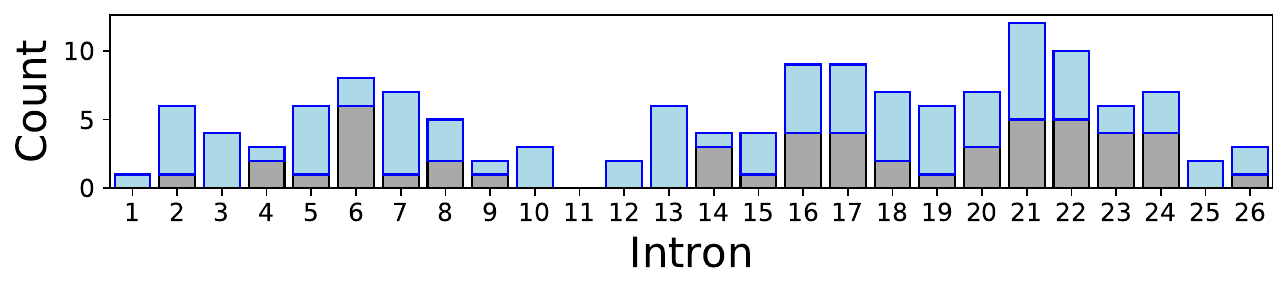}
\caption{BRCA2}
\end{subfigure}
\caption{Distributions of (top) all mutations and (bottom) worst-case pathological mutations across introns of (left) BRCA1 and (right) BRCA2 are provided. In the bottom figures, gray bars represent the amount mutations where best-case and worst-case are both pathological.}
\label{fig:intron_bar_plot_significance}
\end{figure}

\begin{itemize}
    \item \textbf{Gene position}\,\textendash\,Position (of the base pair) in the gene where the mutation occurs, created with position $0$ corresponding to the \texttt{A} of the translation initiation site (\texttt{ATG}).
    \item \textbf{Gene region}\,\textendash\,Describes the region of the mutation as one of the following: exon, intron, other.
    \item \textbf{Detailed gene region}\,\textendash\,Describes the detailed region of the mutation. In the case of exonic or intronic mutations, denotes the exon/intron number of the mutation.
    \item \textbf{Certainty of clinical significance}\,\textendash\,Assesses the level of confidence in the clinical significance of the mutation by categorizing any mutation with \say{unknown} clinical significance as \say{uncertain} and others as \say{certain}.
    \item \textbf{Best-case clinical significance}\,\textendash\,The most benign clinical significance value out of existing ones.
    \item \textbf{Worst-case clinical significance}\,\textendash\,The most pathogenic clinical significance value out of existing ones.
\end{itemize}

Using the aforementioned data with the newly created columns, in what follows, we investigate mutations of BRCA genes.


\section{Results}

To deliver a thorough analysis of BRCA mutations in dbSNP, we adopt a methodical approach, beginning with a broad overview and then gradually delving into more specific details. In what follows, we provide specific details about tables and figures presented in the paper.

\textbf{Genetic structure of BRCA genes}\,\textendash\,Before we explore the mutations in dbSNP, we provide a comprehensive genetic structure (including flanking sequences, exons, and introns) of both BRCA1 and BRCA2 in Figure~\ref{fig:BRCA_Intron_Exon_visulization}. Additionally, we provide the corresponding numbers for each exon directly on top of it. Since the intron lengths are comparably longer than exons, we scale down the intron lengths by $50\%$ in order to improve the visual clarity. Note that the exons and introns are illustrated using different colors (orange for exons and light blue for introns). We will maintain this color palette in the upcoming figures where relevant to enhance clarity. In the later parts of this paper, Figure~\ref{fig:BRCA_Intron_Exon_visulization} will provide valuable context for understanding the mutation locations in these genes.

\begin{figure}[t!]
\centering
\begin{subfigure}{0.49\textwidth}
\includegraphics[width=1\linewidth]{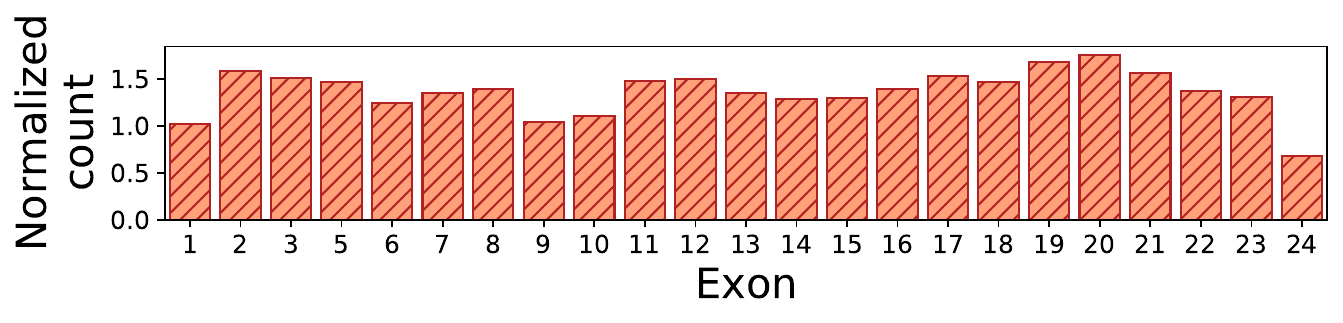}
\includegraphics[width=1\linewidth]{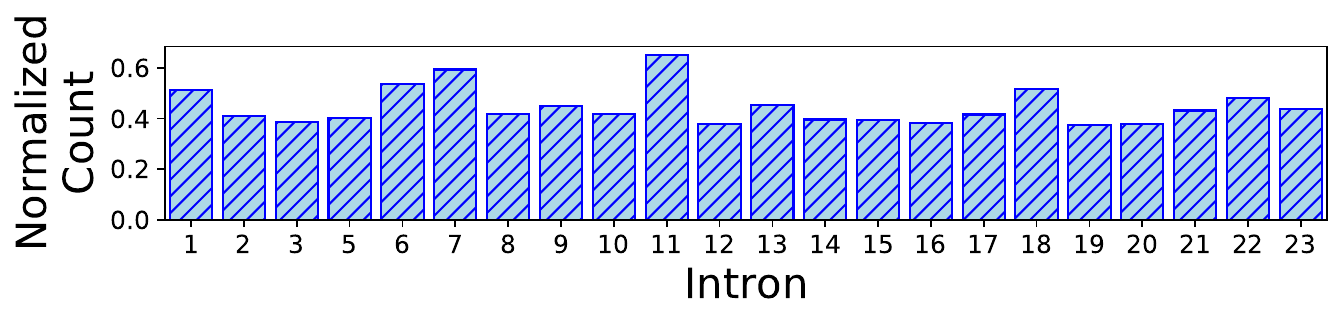}
\caption{BRCA1}
\end{subfigure}
\begin{subfigure}{0.49\textwidth}
\includegraphics[width=1\linewidth]{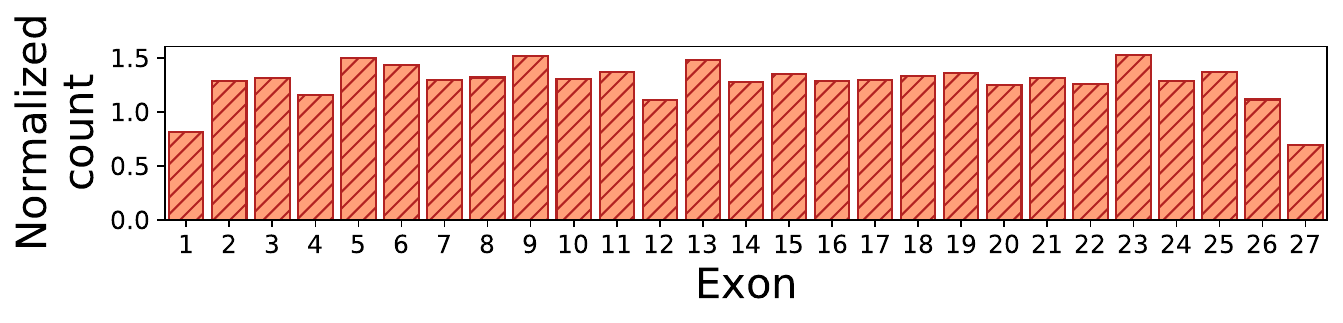}
\includegraphics[width=1\linewidth]{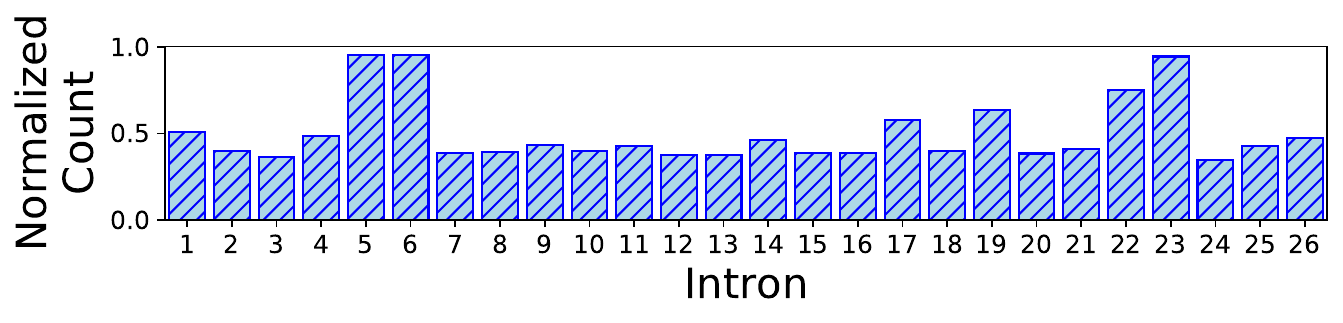}
\caption{BRCA2}
\end{subfigure}
\caption{Mutation counts for (top) exons and (bottom) introns of (left) BRCA1 and (right) BRCA2 are normalized by dividing the total number of observed mutations by their respective exon/intron lengths. This normalization provides a comparative view of mutation frequency, accounting for the variable lengths of exons and introns.}
\label{fig:intron_exon_normalized_total_count}
\end{figure}

\textbf{Overall distribution of mutations}\,\textendash\,In Table~\ref{tbl:total_mutation_count}, we provide the distribution mutations for BRCA1 and BRCA2 across exons, introns, and flanking sequences. For both genes, we observe that the majority of the observed mutations lie in introns ($75\%$ and $64\%$ for BRCA1 and BRCA2, respectively) whereas mutations in exons account for a smaller fraction ($22\%$ for BRCA1 and $33\%$ for BRCA2, respectively). Mutations located in the flanking sequences are relatively rare, making up only approximately 2\% of all mutations in both genes.

\textbf{Mutations across exons and introns}\,\textendash\,Delving deeper into the data presented in Table~\ref{tbl:total_mutation_count}, we provide Figure~\ref{fig:exon_barplot_total_wc_pathogenic_distribution} and Figure~\ref{fig:intron_bar_plot_significance}, which illustrate the distribution of mutations across exons and introns, respectively. In both figures, the bar plots in the first row display data for all mutations, while the ones in the second row present data for mutations with the worst-case clinical significance identified as "pathological." The figures in the second row are further divided into two groups: (gray) mutations where both worst-case and best-case clinical significance are "pathological," and (blue) mutations where the worst-case clinical significance is "pathological," but the best-case clinical significance is different.

It's worth noting that, according to Figure~\ref{fig:exon_barplot_total_wc_pathogenic_distribution}, exon 11 in both BRCA1 and BRCA2 contains the highest number of observed mutations. However, this does not imply that exon 11 is a mutation hotspot. On the contrary, exon 11 is significantly longer than other exons in both genes, which naturally leads to a higher number of mutations due to its length (see Figure~\ref{fig:BRCA_Intron_Exon_visulization} for a comparative view of exons). Recognizing this, we decided to approach the distribution of mutations from a different perspective: investigating the number of observed mutations per base pair, which brings us to the next set of figures.

\textbf{Mutations per base pair}\,\textendash\,In Figure~\ref{fig:intron_exon_normalized_total_count}, we present a modified version of the visualizations shown in the first rows of Figure~\ref{fig:exon_barplot_total_wc_pathogenic_distribution} and Figure~\ref{fig:intron_bar_plot_significance}. Instead of providing the count of mutations per exon/intron, we normalize the mutation count by the length of their respective exon/intron. This normalization allows us to obtain the number of observed mutations per base pair. With Figure~\ref{fig:intron_exon_normalized_total_count}, we validate our earlier observation concerning exon 11, showing that this exon is not particularly unique in terms of mutation distribution. Instead, we find that the majority of exons and introns have a similar number of mutations correlating with their respective lengths.

\textbf{Best- and worst-case clinical significances}\,\textendash\,In Section~\ref{sec:Material and Methods}, we explained how we created two new columns, namely "best-case clinical significance" and "worst-case clinical significance." These new columns contain single clinical significance values, as opposed to the original "clinical significance" column, which could have multiple significance levels. Table~\ref{tbl:BRCA12_clinical_significance_wc_bc_count} shows the distribution of both worst- and best-case clinical significance values across both genes.

Additionally, in Table~\ref{tbl:BRCA12_clinical_significance_wc_bc_count}, we provide a more detailed breakdown of the aforementioned data, specifically focusing on the distribution of worst-case clinical significance values across exons and introns. Note that in both figures, percentages are calculated for each column independently.

\begin{table}[t!] 
\centering
\caption{Mutations in dbSNP for BRCA1 and BRCA2 are categorized according their worst- and best-case clinical significance levels, ranging from ``pathogenic'' to ``benign''.}
\scriptsize
\begin{tabular}{lrrrr}
\cmidrule[0.25pt]{1-5}
~ &  \multicolumn{2}{c}{\shortstack{BRCA1}} & \multicolumn{2}{c}{\shortstack{BRCA2}}\\
\cmidrule[0.25pt]{2-5}
Clinical significance & Worst-case & Best-case & Worst-case & Best-case  \\
\cmidrule[0.5pt]{1-5}
Pathogenic & 2,943 ($7.1\%$) & 2,228 ($5.4\%$) & 3,871 ($8.8\%$) & 2,998 ($6.8\%$) \\
Likely-pathogenic & 428 ($1.0\%$) & 193 ($0.5\%$) & 784 ($1.8\%$) & 414 ($1.0\%$) \\
Unknown & 35,779 ($86.8\%$) & 35,609 ($86.5\%$) & 36,659 ($83.0\%$) & 36,484 ($82.5\%$) \\
Likely-benign & 1,279 ($3.1\%$) & 1,881 ($4.6\%$) & 2,053 ($4.6\%$) & 2,921 ($6.6\%$) \\
Benign & 748 ($1.8\%$) & 1,266 ($3.1\%$) & 838 ($1.9\%$) & 1,388 ($3.1\%$) \\
\cmidrule[1pt]{1-5}
\end{tabular}
\label{tbl:BRCA12_clinical_significance_wc_bc_count}
\end{table}

\begin{table}[t!] 
\centering
\caption{Mutations in dbSNP for BRCA1 and BRCA2 are categorized into exons and introns based on their worst-case clinical significance, ranging from ``pathogenic'' to ``benign''. Percentages are calculated for each category separately.}
\scriptsize
\begin{tabular}{lrrrr}
\cmidrule[0.25pt]{1-5}
~ &  \multicolumn{2}{c}{\shortstack{BRCA1}} & \multicolumn{2}{c}{\shortstack{BRCA2}}\\
\cmidrule[0.25pt]{2-5}
Clinical significance & Exon & Intron & Exon & Intron \\
\cmidrule[0.5pt]{1-5}
Pathogenic & 2,783 ($30.6\%$) & 160 ($0.5\%$) & 3,731 ($25.5\%$) & 139 ($0.5\%$)\\
Likely-pathogenic & 367 ($4.0\%$) & 61 ($0.2\%$) & 689 ($4.7\%$) & 95 ($0.3\%$)\\
Unknown & 5,193 ($57.1\%$) & 29,459 ($95.2\%$) & 8,627 ($59.0\%$) & 27,120 ($94.6\%$)\\
Likely-benign & 722 ($7.9\%$) & 555 ($1.8\%$) & 1531 ($10.5\%$) & 521 ($1.8\%$)\\
Benign & 23 ($0.3\%$) & 701 ($2.3\%$) & 36 ($0.2\%$) & 795 ($2.8\%$)\\
\cmidrule[1pt]{1-5}
\end{tabular}
\label{tbl:BRCA_intron_exon_clinical_significance}
\end{table}

\textbf{Mutation variants}\,\textendash\,Thus far we have not made a distinction across mutation variant types. In Table~\ref{tbl:BRCA_both_worst_case_significance_variant_type_analysis}, we provide the distribution of mutations for each variant type for various levels of worst-case clinical significance. For the variant type column, we use the values directly from dbSNP (variant type descriptions are provided in Section~\ref{sec:Material and Methods}).

\begin{table}[t!] 
\centering
\scriptsize
\caption{Mutations in dbSNP for BRCA1 and BRCA2 are categorized based on their variant type (del, delins, ins, mnv, snv) into worst-case clinical significance, ranging from ``pathogenic'' to ``benign''.}
\begin{tabular}{llrrrrr}
\cmidrule[0.25pt]{1-7}
Gene & Variant type & Benign & Likely-benign & Unknown & Likely-pathogenic& Pathogenic \\
\cmidrule[0.5pt]{1-7}
\multirow{5}{*}{\rotatebox[origin=c]{0}{BRCA1}} & del & 14 ($1.9\%$) & 14 ($1.1\%$) & 834 ($2.3\%$) & 19 ($4.4\%$) & 613 ($20.8\%$) \\
~ & delins & 235 ($31.4\%$) & 113 ($8.8\%$) & 2,998 ($8.4\%$) & 83 ($19.4\%$) & 1,468 ($49.9\%$) \\
~ & ins & 6 ($0.8\%$) & 5 ($0.4\%$) & 395 ($1.1\%$) & 3 ($0.7\%$) & 210 ($7.1\%$) \\
~ & mnv & 0 ($0.0\%$) & 7 ($0.5\%$) & 30 ($0.1\%$) & 2 ($0.5\%$) & 9 ($0.3\%$) \\
~ & snv & 493 ($65.9\%$) & 1,140 ($89.1\%$) & 31,522 ($88.1\%$) & 321 ($75.0\%$) & 643 ($21.8\%$) \\
\cmidrule[0.5pt]{1-7}
\multirow{5}{*}{\rotatebox[origin=c]{0}{BRCA2}} & del & 20 ($2.4\%$) & 22 ($1.1\%$) & 761 ($2.1\%$) & 37 ($4.7\%$) & 692 ($17.9\%$) \\
~ & delins & 222 ($26.5\%$) & 92 ($4.5\%$) & 2,640 ($7.2\%$) & 140 ($17.9\%$) & 2,148 ($55.5\%$) \\
~ & ins & 8 ($1.0\%$) & 6 ($0.3\%$) & 310 ($0.8\%$) & 5 ($0.6\%$) & 191 ($4.9
\%$) \\
~ & mnv & 0 ($0.0\%$) & 5 ($0.2\%$) & 58 ($0.2\%$) & 5 ($0.6\%$) & 17 ($0.4\%$) \\
~ & snv & 588 ($70.2\%$) & 1,928 ($93.9\%$) & 32,890 ($89.7\%$) & 597 ($76.1\%$) & 823 ($21.3\%$) \\
\cmidrule[1pt]{1-7}
\end{tabular}
\begin{tabular}{llccccc}
\cmidrule[1pt]{1-6}
\end{tabular}
\label{tbl:BRCA_both_worst_case_significance_variant_type_analysis}
\end{table}

\textbf{Certainty of clinical significance}\,\textendash\,As mentioned in Section~\ref{sec:Material and Methods}, we determined the certainty of clinical significance based on the existing values of clinical significance for each observed mutation. Table~\ref{tbl:BRCA_intron_exon_clinical_significance} displays the distribution of certainty of observations for both genes across exons and introns. Moreover, in Table~\ref{tbl:BRCA_intron_exon_clinical_significance}, we specifically present the certainty of a subset of mutations where the worst-case clinical significance is labeled as ``pathogenic.'' To complement this table, we visually represent the certainty of clinical significance for mutations in exons and introns in Figure~\ref{fig:exon_barplot_wc_pathogenic_normalized_certainty}.

\begin{table}[t!] 
\centering
\scriptsize
\caption{All mutations in dbSNP for BRCA1 and BRCA2, as well as only the ones that are worst-case pathogenic, are categorized into exons and introns based on the certainty of their clinical significance (see Section-X for further details).}
\begin{tabular}{llrrrr}
\cmidrule[0.25pt]{1-6}
~ & ~ & \multicolumn{2}{c}{\shortstack{All Mutations}} & \multicolumn{2}{c}{\shortstack{Pathogenic}} \\
\cmidrule[0.25pt]{3-6}
Gene & Certainty & Exon & Intron & Exon & Intron \\
\cmidrule[0.5pt]{1-6}
\multirow{2}{*}{\rotatebox[origin=c]{0}{BRCA1}} & Certain & 3,126 ($34.4\%$) & 1,371 ($4.4\%$) & 2,275 ($81.8\%$) & 90 ($56.3\%$)\\
~ & Uncertain & 5,962 ($65.6\%$) & 29,565 ($95.6\%$) & 508 ($18.3\%$) & 70 ($43.8\%$)\\
\cmidrule[0.5pt]{1-6}
\multirow{2}{*}{\rotatebox[origin=c]{0}{BRCA2}} & Certain & 4,943 ($33.8\%$) & 1,485 ($5.2\%$) & 3,185 ($85.4\%$) & 113 ($81.3\%$)\\
~ & Uncertain & 9,671 ($66.2\%$) & 27,185 ($94.8\%$) & 546 ($14.6\%$) & 26 ($18.7\%$)\\
\cmidrule[1pt]{1-6}
\end{tabular}
\label{tbl:BRCA_both_certainty_table}
\end{table}

\textbf{Distribution of mutations within exons and introns}\,\textendash\,Up to this point, when we categorized mutations into exons and introns, we haven't considered their specific positions within these regions. This task is challenging since each exon and intron varies in length (see Figure~\ref{fig:BRCA_Intron_Exon_visulization}). To create a visualization that captures the relative positions of mutations within exons and introns, we preprocess the data by calculating their relative positions in percentage intervals of $5\%$. For instance, if an exon has a length of $2,000$ base pairs, all mutations falling between positions $1$ and $100$ would be placed in the $0-5$ percentage interval. This approach enables us to explore mutations in terms of their relative positions and allows us to make relevant observations regarding their distribution within exons and introns.

By implementing the aforementioned approach, we present Figure~\ref{fig:BRCA_percent_interval_tot}, which provides visualizations of mutations within exons (orange) and introns (blue) divided into $5\%$ intervals. Figure~\ref{fig:percentage_all_muts} displays the obtained results for all mutations, while Figure~\ref{fig:percentage_pathogenic} focuses specifically on mutations with a pathogenic worst-case clinical significance. 



\section{Discussion}
\label{sec:Discussion}
We now briefly explain the ramifications of our experimental observations.

\textbf{Distribution of mutations across exons and introns}\,\textendash\,Figure~\ref{fig:BRCA_Intron_Exon_visulization} visually illustrates the structure of BRCA1 and BRCA2 based on their exon and intron lengths. In BRCA1, introns span a total length of 73,982 bp, which is 10.4 times longer than the cumulative length of exons, comprising 7,088 bp. Similarly in BRCA2, introns cover 72,807 bp, exhibiting a length 6.1 times greater than the 11,954 bp of exons~\cite{10.1093/nar/gkab1049}. Table~\ref{tbl:total_mutation_count} presents the comprehensive count of mutations in BRCA1 and BRCA2, retrieved from the dbSNP and categorized based on their respective locations within each gene. Our analysis reveals a notable difference, indicating that introns exhibit 3.4 and 2.0 times more mutations compared to exons in BRCA1 and BRCA2 respectively.

Given the difference in ratios between segment length and mutation count, we conducted further exploration of mutation frequency based on the exon/intron length. For BRCA1, the occurrence of mutations per bp was 0.39 in introns and 1.22 in exons, indicating a higher frequency of mutations in exonic regions compared to intronic regions on a per-base pair basis. Similarly for BRCA2, these values are 0.44 and 1.22 for introns and exons respectively, implying a similar trend of higher mutation frequency in exonic regions.

\begin{figure}[t!]
\centering
\begin{subfigure}{0.49\textwidth}
\includegraphics[width=1\linewidth]{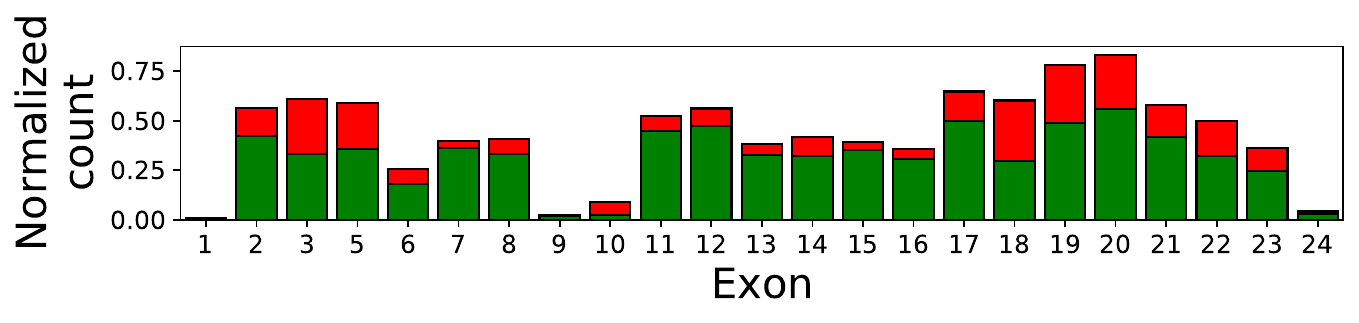}
\includegraphics[width=1\linewidth]{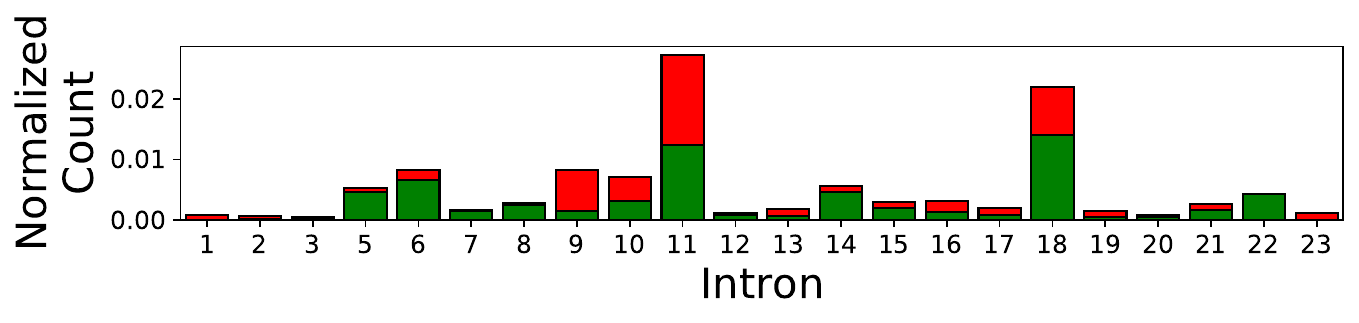}
\caption{BRCA1}
\end{subfigure}\begin{subfigure}{0.49\textwidth}
\includegraphics[width=1\linewidth]{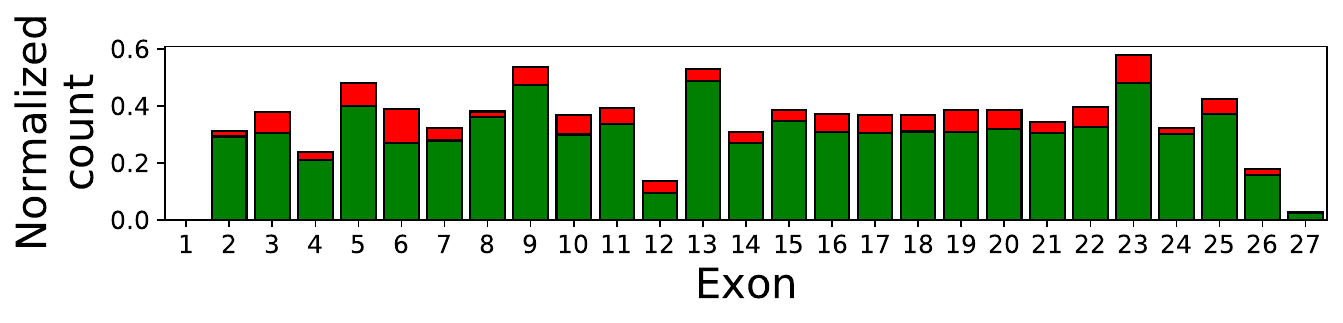}
\includegraphics[width=1\linewidth]{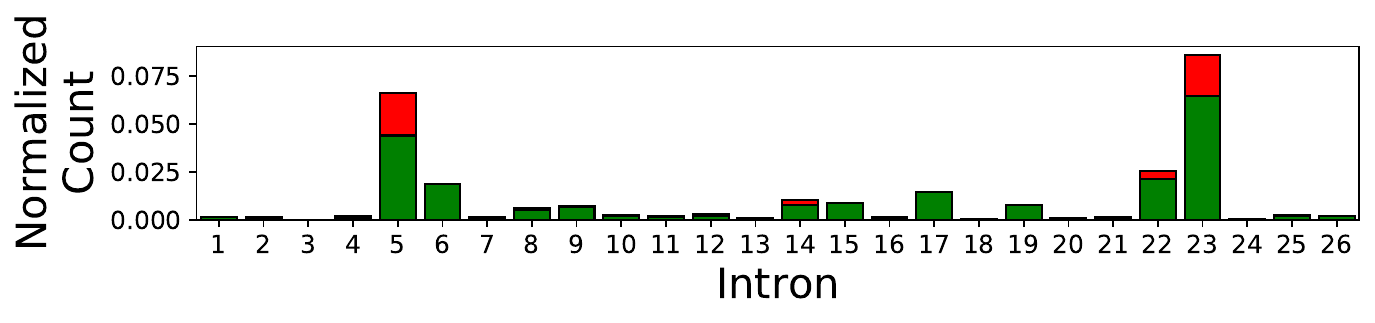}
\caption{BRCA2}
\end{subfigure}
\caption{Distribution of genetic mutations with certain (visualized with green) and uncertain (visualized with red) clinical significances for (left) BRCA1 and (right) BRCA2 is presented for their respective (top) exons and (bottom) introns.}
\label{fig:exon_barplot_wc_pathogenic_normalized_certainty}
\end{figure}

\textbf{Pathogenicity of mutations in exons and introns}\,\textendash\, 
According to Table~\ref{tbl:BRCA_intron_exon_clinical_significance}, more than $90\%$ of the mutations in introns have an unknown clinical significance, whereas, $57.1\%$ and $59.0\%$ of exonic mutations in BRCA1 and BRCA2, respectively have an unknown clinical significance. The occurrence of pathogenicities other than `unknown' in introns for both BRCA genes is low, with a frequency of less than $5\%$. In exons, the frequency of pathogenic mutations is comparatively high, at $30.6\%$ and $25.5\%$ for BRCA1 and BRCA2, respectively. 



\textbf{Excessive number of mutations in exon 11 of BRCA1 and BRCA2}\,\textendash\,The distribution of mutations across introns and exons is illustrated in Figure~\ref{fig:exon_barplot_total_wc_pathogenic_distribution}. Notably, BRCA1 and BRCA2 exhibit a high mutation frequency in exon 11. This exon contains splicing regulatory sequences at its beginning, resulting in the generation of multiple splicing isoforms~\cite{tammaro2012brca1}. The accumulation of these isoforms potentially contributes to altered splicing profiles of the gene, affecting its essential function in DNA repair~\cite{tammaro2012brca1}. Additionally, Figure~\ref{fig:intron_exon_normalized_total_count}, displaying the normalized count of mutations for each exon, reveals that exons tend to have approximately the same number of mutations as their respective base pair count. Exon 11 of BRCA1 and BRCA2 spans 3,426 and 4,932 bps respectively, making it one of the largest human exons involved in DNA repair, cell growth, and cell cycle control~\cite{raponi2014brca1}. Our analysis reveals a relatively even distribution of exonic mutations when normalized for length, indicating that mutations are spread out uniformly within the exonic regions. Therefore, according to Figure~\ref{fig:exon_barplot_total_wc_pathogenic_distribution} and Figure~\ref{fig:intron_exon_normalized_total_count}, we can say that the high abundance of mutations in exon 11 largely arises due to its length rather than a specific mutation hotspot.

\textbf{Number of unknown mutations}\,\textendash\,As depicted in Table~\ref{tbl:BRCA12_clinical_significance_wc_bc_count}, the clinical significance of over $80\%$ of mutations remains unknown, indicating an unclear impact of these genetic alterations. Mutations can exhibit a wide range of effects on the function of genes. Some mutations may have uncertain impacts, posing challenges in their interpretation, and these variants might be listed with varying clinical significances. During the data processing step, mutations classified as `not-provided,' `uncertain-significance', `other', `unknown', or containing null values were collectively redefined as `unknown'. With this approach, our goal was to address multiple instances of unclear clinical significance and to unify them. 

With the increasing number of genetic testings being conducted, more variants of uncertain significance (VUS) are identified. These variants lack sufficient evidence to be classified as pathogenic mutations~\cite{cheon2014variants}. In the case of BRCA, a VUS result does not provide a clear indication of whether the patient is at a higher risk of developing breast cancer. Therefore, methods and guidelines have been proposed to reevaluate and reclassify VUS as more comprehensive information becomes available~\cite{huszno2021brca1}. 

\begin{figure}[t!]
\centering
\begin{subfigure}{0.98\textwidth}
\includegraphics[width=0.46\linewidth]{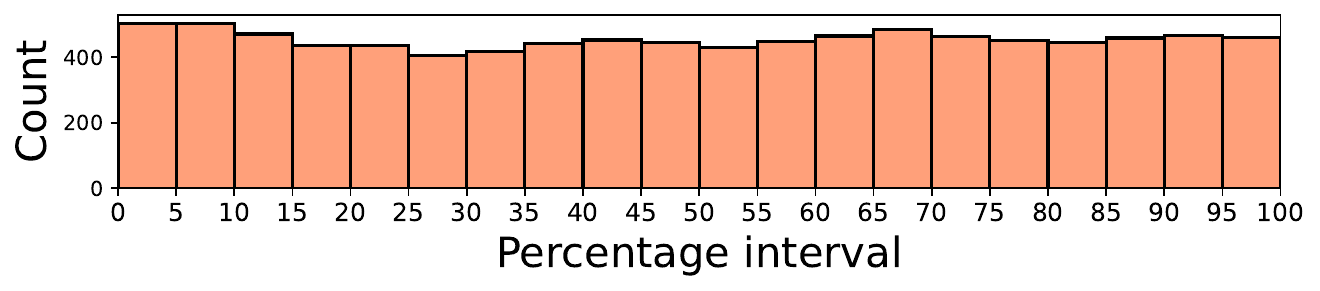}
\includegraphics[width=0.46\linewidth]{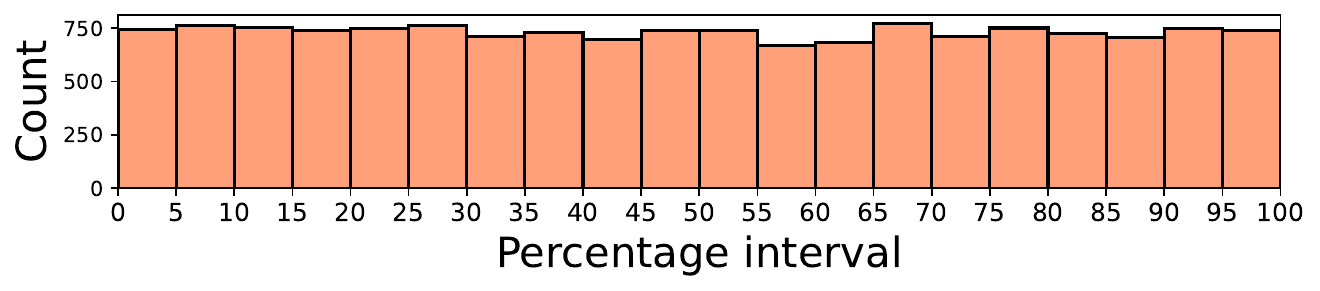}
\\
\includegraphics[width=0.46\linewidth]{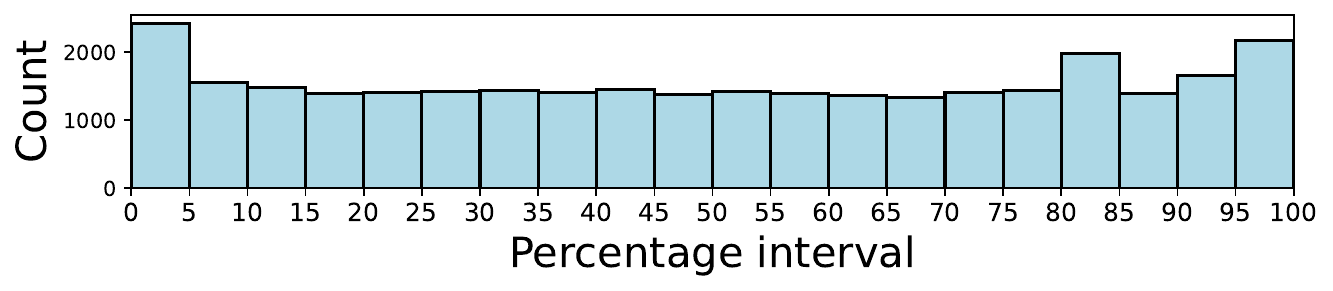}
\includegraphics[width=0.46\linewidth]{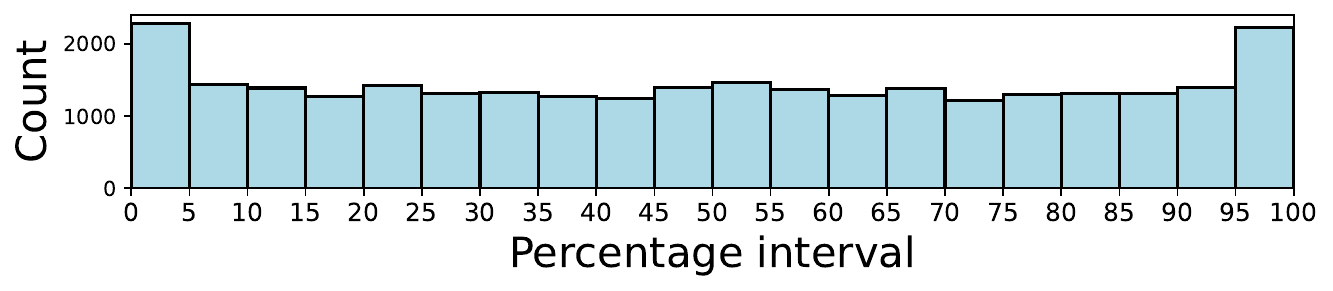}
\caption{All mutations}
\label{fig:percentage_all_muts}
\end{subfigure}
\\
\phantom{-}
\\
\begin{subfigure}{0.98\textwidth}
\includegraphics[width=0.46\linewidth]{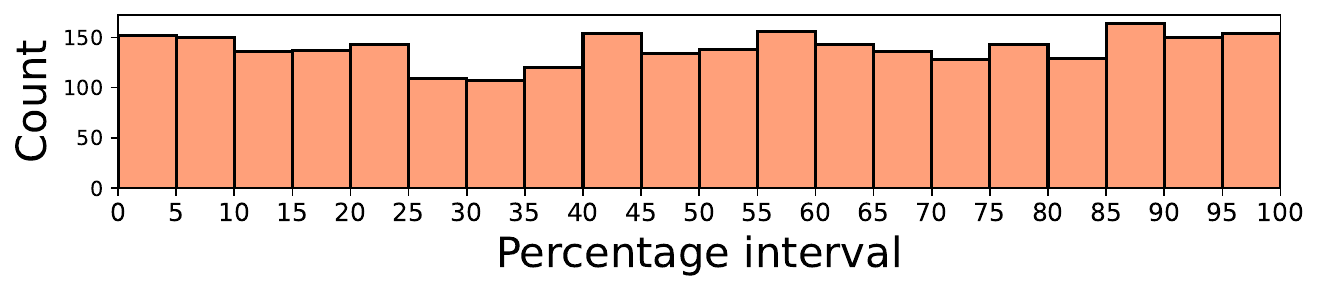}
\includegraphics[width=0.46\linewidth]{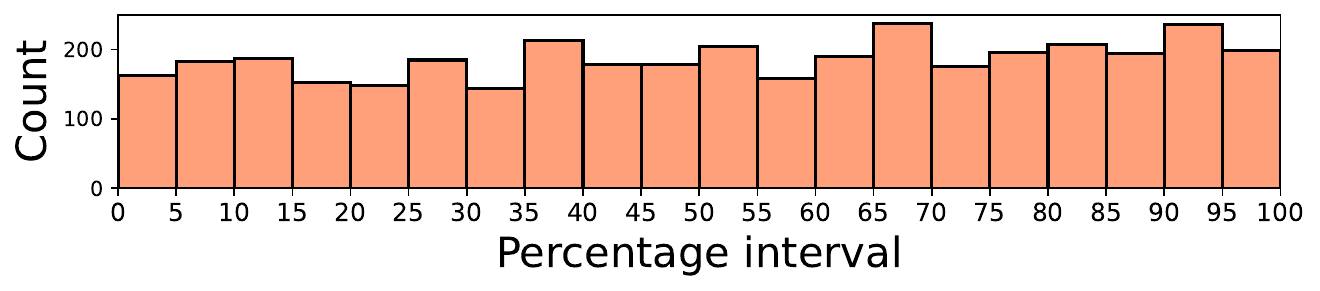}
\\
\includegraphics[width=0.46\linewidth]{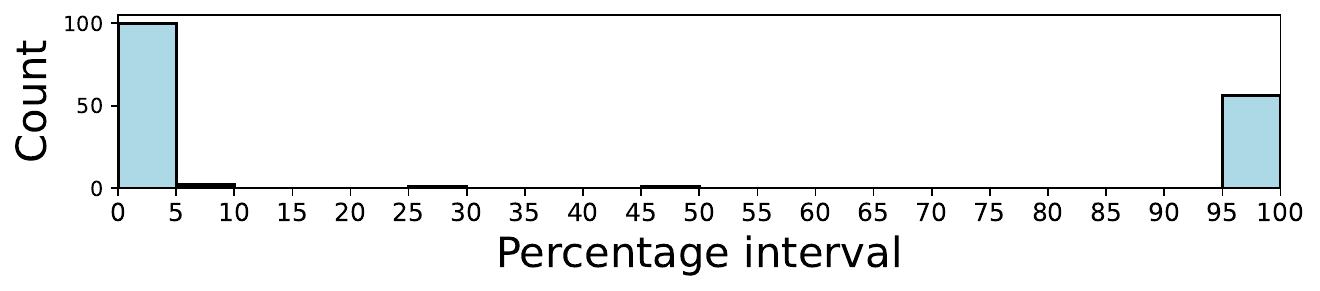}
\includegraphics[width=0.46\linewidth]{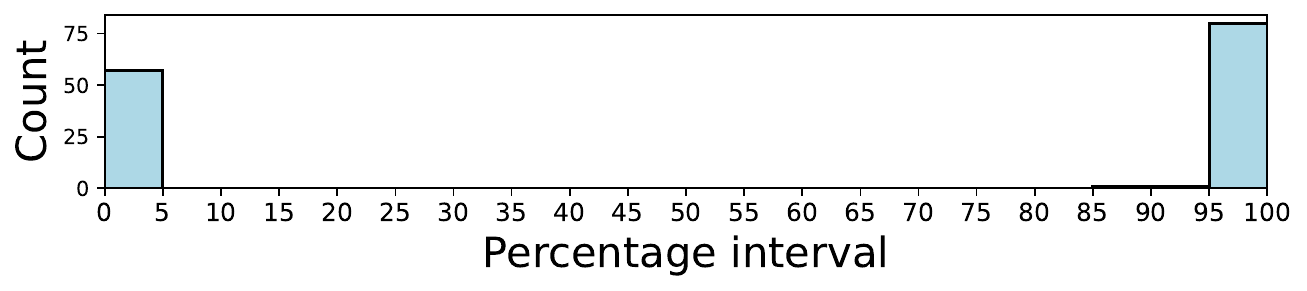}
\caption{Pathogenic mutations}
\label{fig:percentage_pathogenic}
\end{subfigure}
\caption{Mutation count in (\textbf{left}) BRCA1 and (\textbf{right})  BRCA2 based on percentage intervals. Histograms display the total number of mutations in (\textbf{left}) BRCA1 and (\textbf{right}) BRCA2 in different regions of exons or introns. Each exon and intron is divided into $5\%$ intervals based on their length. Mutations within each interval region are counted to obtain the total mutation count for all exons or introns in their respected regions. The first bar in each histogram represents the number of mutation incidences within the first $5\%$ region, while the last bar corresponds to the $95\%$ to $100\%$ region. The top histograms show the distribution of mutations in exons, and the bottom histograms show the distribution of mutations in introns.}
\label{fig:BRCA_percent_interval_tot}
\end{figure}

\textbf{Certainty of observations}\,\textendash\,Mutations that are ‘Certain’ are likely to have clear clinical implications, and are known to have documented functional consequences on the encoded protein. The conflicting certainty regarding pathogenicity may be attributed to differences in the functions and interactions of BRCA1 and BRCA2 genes within the cells, as well as variations in the types and locations of mutations commonly observed in each gene~\cite{petrucelli2010hereditary}. These factors can influence the clinical impact of the mutations and the level of certainty in defining their pathogenicity. According to Table~\ref{tbl:BRCA_both_certainty_table}, we observe that exonic mutations that are worst-case pathogenic are mostly certain, indicating that their effects are documented. Specifically, there is a higher level of certainty in identifying worst-case pathogenic mutations in BRCA2, as shown in Figure ~\ref{fig:exon_barplot_wc_pathogenic_normalized_certainty}. 

Although mutations in BRCA1 are associated with higher cancer risk, with a $39\%$ risk of developing cancer by the age of 70, compared to an $11\%$ risk with BRCA2 pathogenic mutations, there are more documented pathogenic variants in BRCA2~\cite{albert2020brca1}. Supporting our observations,~\cite{albert2020brca1} also reports that BRCA1 has approximately 1,600 documented pathogenic variants whereas BRCA2 has approximately 1,800.

\textbf{Mutation variant types}\,\textendash\,In Table~\ref{tbl:BRCA_both_worst_case_significance_variant_type_analysis}, most of the mutations in dbSNP are "Single nucleotide variant" (SNV). Despite the prevalence of SNV, their clinical significance remains largely unknown ($88.1\%$ and $89.7\%$ of snv mutations in BRCA1 and BRCA2). The mutation variant type "deletion" (del) and "insertion" (ins) both exhibit high percentage of pathogenic cases in BRCA1 and BRCA2. The mutation variants in BRCA1 and BRCA2 that introduce frameshifts often result in the production of missense or non-functional proteins~\cite{karami2013comprehensive}. Insertion and deletion, which may cause a shift in the open reading frame, result in the formation of premature stop codons, leading to the lack of a properly functioning protein involved in DNA repair, and ultimately affecting the prevention of cancer~\cite{mehrgou2016importance}~\cite{abul2020exome}.

\textbf{Mutations in splicing regions}\,\textendash\,Patterns for pathogenic intronic mutations are observed through Figure~\ref{fig:percentage_pathogenic}. In contrast to pathogenic exonic mutations, which tend to exhibit a relatively even distribution due to their significant impact on translated proteins, pathogenic intronic mutations clustered predominantly at the edges. The edges of introns are where highly conserved sequences called splice site regions are located, and are critical in facilitating accurate splicesome recognition. Any mutations within these critical regions may disrupt the splicing process, resulting in aberrant mRNA maturation\cite{rogan1998information}.

Splice site mutations contributes for $3.9\%$ and $2.2\%$ of total mutations in BRCA1 and BRCA2 respectively\cite{lopez2019brca}. In our case, Figure~\ref{fig:percentage_pathogenic} indicates that $99\%$ of all pathogenic mutations in introns are either at the beginning (0-5\% interval) or at the end (95\% to 100\% interval). The clustering of pathogenic mutations within splice site regions sheds light on their functional importance in accurate splicing. Disruption of the splicing process due to mutations in splice site regions causes incorrect exon inclusion or essential exon exclusion, resulting in the production of dysfunctional proteins. As such, understanding the molecular mechanisms underlying these effects is crucial for deciphering the basis of genetic conditions and exploring potential therapeutic interventions.

\section{Conclusions and Future Perspectives}

In this work, we presented a comprehensive analysis of BRCA1 and BRCA2 mutations in dbSNP. By incorporating effective data summarization as well as visualization methods, we thrived to create a clear and comprehensive view of the mutation patterns within these genes. Our analysis revealed a notable correlation between mutation frequency and the length of exons, with a higher number of mutations occurring on a per-base pair basis in the exonic regions. Additionally, we observed that within introns, pathogenic mutations tend to cluster around the splice site regions at the beginning and end. 


In Section~\ref{sec:Discussion}, we denoted that there are more mutations in exons than in introns on a per bp basis in dbSNP. The higher prevalence of mutations in exons can be attributed to their greater significance in genetic studies, as these regions often harbor critical functional elements. However, it is essential to acknowledge that our analysis primarily focused on the total number of mutation recorded in the dbSNP, neglecting consideration of the frequency of the mutations. To gain a more comprehensive insight into the mutation patterns of BRCA1 and BRCA2 and their prevalence in the population, further research incorporating the aforementioned data is necessary.

Our analysis of pathogenic mutations based on percentage intervals has highlighted the functional significance of splice site sequences. The presence of pathogenic mutations within these critical splice site regions underscores their crucial role in facilitating accurate splicing and mRNA maturation. Disruptions to the splicing process caused by mutations in these regions can lead to the production of aberrant proteins, contributing to the development of genetic conditions. Therefore, understanding the molecular mechanisms underlying the effects of these splice site mutations is of utmost significance. However, it is crucial to recognize specific limitations in our approach. The inclusion of all mutations within BRCA1 and BRCA2 in the analysis might have obscured potential differences in mutation frequencies within specific introns. To gain a more precise understanding of potentially fragile genomic regions, further investigation focusing on individual introns is needed.

\bibliographystyle{unsrtnat}
\bibliography{brca_main}
\end{document}